\input phyzzx
\overfullrule=0pt
\def\tr{\mathop{\rm tr}}
\def\Tr{\mathop{\rm Tr}}
%
\REF\GIN{%
P. H. Ginsparg and K. G. Wilson,
Phys.\ Rev.\ {\bf D25} (1982) 2649.}
\REF\HAS{%
P. Hasenfratz,
Nucl.\ Phys.\ (Proc.\ Suppl.) {\bf 63} (1998) 53;
Nucl.\ Phys.\ {\bf B525} (1998) 401.}
\REF\HASS{%
P. Hasenfratz, V. Laliena and F. Niedermayer,
Phys.\ Lett.\ {\bf B427} (1998) 125.}
\REF\NEU{%
H. Neuberger,
Phys.\ Lett.\ {\bf B417} (1998) 141; {\bf B427} (1998) 353.}
\REF\LUS{%
M. L\"uscher,
Phys.\ Lett.\ {\bf B428} (1998) 342.}
\REF\FUJ{%
K. Fujikawa,
Phys.\ Rev.\ Lett.\ {\bf 42} (1979) 1195; Phys.\ Rev.\ {\bf D21}
(1980) 2848; {\bf D22} (1980) 1499~(E); {\bf D29} (1984) 285.}
\REF\KIK{%
Y. Kikukawa and A. Yamada,
Phys.\ Lett.\ {\bf B448} (1999) 265.}
\REF\ADA{%
D. H. Adams,
hep-lat/9812003.\nextline
H. Suzuki,
Prog.\ Theor.\ Phys.\ {\bf 102} (1999) 141.}
\REF\FUJI{%
K. Fujikawa,
Nucl.\ Phys.\ {\bf B546} (1999) 480.}
\REF\CHI{%
T.-W. Chiu,
Phys.\ Lett.\ {\bf B445} (1999) 371.\nextline
T.-W. Chiu and T.-H. Hsieh,
hep-lat/9901011.\nextline
T. Reisz and H. J. Rothe,
Phys.\ Lett.\ {\bf B455} (1999) 246.}
\REF\LUSC{%
M. L\"uscher,
Nucl.\ Phys.\ {\bf B538} (1999) 515.}
\REF\NAR{%
R. Narayanan,
Phys.\ Rev.\ {\bf D58} (1998) 097501.\nextline
F. Niedermayer,
Nucl.\ Phys.\ (Proc.\ Suppl.) {\bf 73} (1999) 105.\nextline
Y. Kikukawa and A. Yamada,
Nucl.\ Phys.\ {\bf B547} (1999) 413.}
\REF\NEUB{%
H. Neuberger,
Phys.\ Rev.\ {\bf D59} (1999) 085006.}
\REF\NARA{%
R. Narayanan and H. Neuberger,
Nucl.\ Phys.\ {\bf B412} (1994) 574; {\bf B443} (1995) 305.\nextline
S. Randjbar-Daemi and J. Strathdee,
Phys.\ Lett.\ {\bf B402} (1997) 134.}
\REF\LUSCH{%
M. L\"uscher,
Nucl.\ Phys.\ {\bf B549} (1999) 295.}
\REF\FUJIK{%
K. Fujikawa,
Phys.\ Rev.\ {\bf D60} (1999) 074505.\nextline
T. Aoyama and Y. Kikukawa,
hep-lat/9905003.\nextline
T.-W. Chiu,
hep-lat/9906007.}
\REF\BRA{%
F. Brandt, N. Dragon and M. Kreuzer,
Phys.\ Lett.\ {\bf B231} (1989) 263; Nucl.\ Phys.\ {\bf B332} (1990)
224; {\bf B332} (1990) 250.\nextline
N. Dragon,
Lectures given at Saalburg Summer School (1995), hep-th/9602163.}  
\REF\DUB{%
M. Dubois-Violette, M. Henneaux, M. Talon and C.-M. Viallet,
Phys.\ Lett.\ {\bf B267} (1991) 81; {\bf B289} (1992) 361.}
\REF\ZUM{%
B. Zumino, Y. S. Wu and A. Zee,
Nucl.\ Phys.\ {\bf B239} (1984) 477.\nextline
B. Zumino,
in {\sl Relativity, Groups and Topology II}, eds.\ B. S. De Witt and
R. Stora, (North-Holland, Amsterdam, 1984).\nextline
L. Baulieu,
Nucl.\ Phys.\ {\bf B241} (1984) 557;
in {\sl Progress in Gauge Field Theory}, eds.\ H. Lehmann et.\ al.,
NATO ASI Series B, Physics, Vol.~115 (Plenum, New York,
1984).\nextline
R. Stora,
in {\sl Progress in Gauge Field Theory}, eds.\ H. Lehmann et.\ al.,
NATO ASI Series B, Physics, Vol.~115 (Plenum, New York, 1984).}
\REF\CON{%
A. Connes, {\sl Noncommutative Geometry}, (Academic Press, New York,
1994).}
\REF\SIT{%
A. Sitarz,
J. Geom.\ Phys.\ {\bf 15} (1995) 123.}
\REF\DIM{%
A. Dimakis and F. M\"uller-Hoissen,
Phys.\ Lett.\ {\bf B295} (1992) 242;
J. Phys. A: Math.\ Gen.\ {\bf 27} (1994) 3159;
J.\ Math.\ Phys.\ {\bf 35} (1994) 6703.\nextline
A. Dimakis, F. M\"uller-Hoissen and T. Striker, 
Phys.\ Lett.\ {\bf B300} (1993) 141;
J. Phys. A: Math.\ Gen.\ {\bf 26} (1993) 1927.}
\REF\DIN{%
H. G. Ding, H. Y. Guo, J. M. Li and K. Wu,
Z. Phys.\ {\bf C64} (1994) 521; J. Phys.\ A: Math.\ Gen.\ {\bf 27}
(1994) L75; {\bf 27} (1994) L231; Commun.\ Theor.\ Phys.\ {\bf 21}
(1994) 85.\nextline
H. Y. Guo, K. Wu and W. Zhang,
``Noncommutative Differential Calculus on Discrete Abelian Groups and
Its Applications,'' ITP-Preprint, March, 1999.}
\REF\BAL{%
A. P. Balachandran, G. Bimonte, G. Landi, F. Lizzi and
P. Teotonio-Sobrinho,
J. Geom.\ Phys.\ {\bf 24} (1998) 353, and references therein.}
\REF\FUJIW{%
T. Fujiwara, H. Suzuki and K. Wu,
Phys.\ Lett.\ {\bf B463} (1999) 63.}
\REF\LUSCHE{%
M. L\"uscher,
hep-lat/9904009.}
\REF\HOR{%
I. Horvath,
Phys.\ Rev.\ Lett.\ {\bf 81} (1998) 4063;
Phys.\ Rev.\ {\bf D60} (1999) 034510.\nextline
W. Bietenholz, hep-lat/9901005.}
\REF\HER{%
P. Hern\'andez, K. Jansen and M. L\"uscher,
Nucl.\ Phys.\ {\bf B552} (1999) 363.}
\REF\DIMA{%
A. Dimakis and F. M\"uller-Hoissen,
physics/9712004.}
\REF\SUZ{%
H. Suzuki,
Prog.\ Theor.\ Phys.\ {\bf 101} (1999) 1147.}
%
\Pubnum={IU-MSTP/35\cr hep-lat/9906015}
\date={June 1999}
\titlepage
\title{Non-commutative Differential Calculus and\break
the Axial Anomaly in Abelian Lattice Gauge Theories}
\author{%
Takanori Fujiwara,\foot{%
E-mail: fujiwara@mito.ipc.ibaraki.ac.jp}
Hiroshi Suzuki\foot{%
E-mail: hsuzuki@mito.ipc.ibaraki.ac.jp}
and Ke Wu\foot{%
On leave of absence from Institute of Theoretical Physics, Academia
Sinica, P. O. Box 2735, Beijing 100080, China.}
\foot{%
E-mail: wuke@mito.ipc.ibaraki.ac.jp}}
\address{%
Department of Mathematical Sciences, Ibaraki University, Mito
310-8512, Japan}
\abstract{%
The axial anomaly in lattice gauge theories has a topological nature
when the Dirac operator satisfies the Ginsparg-Wilson relation. We
study the axial anomaly in Abelian gauge theories on an infinite
hypercubic lattice by utilizing cohomological arguments. The crucial
tool in our approach is the non-commutative differential
calculus~(NCDC) which makes the Leibniz rule of exterior
derivatives valid on the lattice. The topological nature of the
``Chern character'' on the lattice becomes manifest in the context of
NCDC. Our result provides an algebraic proof of L\"uscher's theorem
for a four-dimensional lattice and its generalization to arbitrary
dimensions.}
\bigskip\medskip
\noindent
PACS numbers: 11.15.Ha, 11.30.Rd, 02.40.-k\hfill\break
\endpage
\chapter{Introduction}
The non-perturbative formulation of chiral gauge theories is a long
standing problem in theoretical particle physics. It appears that, if
one wishes to regularize chiral gauge theories by keeping the gauge
symmetry manifest, one has to first understand the structure of gauge
anomalies defined in a theory with a finite ultraviolet cutoff.
Therefore it is quite important to study the axial (gauge) anomalies
on the lattice while keeping the lattice spacing finite. 

Recently, there have been remarkable developments regarding this
problem. First, gauge covariant local Dirac operators which satisfy
the Ginsparg-Wilson~(GW) relation~[\GIN] were
discovered~[\HAS,\HASS,\NEU]. It was soon recognized that GW relation
implies the index theorem~[\HASS,\LUS] and the exact chiral
symmetry~[\LUS] of the fermion action. The chiral symmetry is however
anomalous due to the non-trivial Jacobian factor and the Jacobian is
related to the index, as is in the continuum theory~[\FUJ]. The
Jacobian associated with the local chiral transformation, i.e., the
chiral or axial anomaly, is given by the GW Dirac operator~$D$
as\foot{%
Throughout this article, the lattice spacing is taken to be
unity~$a=1$.}
$$
   q(x)=\tr\gamma_5\left[\delta_{x,x}-{1\over2}D(x,x)\right],
\eqn\onexone
$$
where $\tr$ stands for the trace over the spinor indices. The
perturbative evaluation of the quantity~$q(x)$ in the continuum limit
was carried out in~Ref.~[\KIK] by using the overlap-Dirac
operator~[\NEU] and it was confirmed that the field $q(x)$~reproduces
the anomaly in the continuum theory. See also~Ref.~[\ADA]. It can
even be shown~[\FUJI] that the continuum limit of~Eq.~\onexone\ in
general coincides with the axial anomaly in the continuum theory, if
the Dirac operator behaves properly in this limit. This issue is
addressed also in~Ref.~[\CHI].

The index theorem~[\HASS,\LUS] states that the volume integral of the
axial anomaly~$q(x)$ is integer valued. This would imply that the
volume integral of~$q(x)$ is invariant under a local variation of the
gauge field\foot{%
This topological property of the anomaly can be shown even on an
infinite lattice, for which notion of the index might be ill-defined.
Introduce the operator~${\mit\Gamma}_5=\gamma_5(1-D)$. Then GW
relation~$\gamma_5D+D\gamma_5=D\gamma_5D$ implies
${\mit\Gamma}_5^2=1$ and~$\{{\mit\Gamma_5},\gamma_5\delta D\}=0$.
>From these, we have $\Tr\gamma_5\delta D=%
\Tr\gamma_5\delta D{\mit\Gamma}_5^2=%
\Tr{\mit\Gamma}_5\gamma_5\delta D{\mit\Gamma}_5=%
- -\Tr\gamma_5\delta D=0$, here $\Tr$~stands for $x$-integration of the
diagonal ($xx$) components as well as the spinor trace. This is
equivalent to~$\sum_x\delta q(x)=0$. Note that the
sum~$\Tr\gamma_5\delta D$ is well-defined even on the infinite
lattice, at least if the Dirac operator~$D$ has a local dependence on
the gauge potential and if the variation of the gauge potential is
well-localized.}
$$
   \sum_x\delta q(x)=0.
\eqn\onextwo
$$
Solely from this topological property combined with the gauge
invariance and the locality of the field~$q(x)$, it was shown~[\LUSC]
that, in Abelian gauge theories on a four-dimensional infinite
hypercubic lattice, the field $q(x)$~has the following structure:
$$
   q(x)=\alpha+\beta_{\mu\nu}F_{\mu\nu}(x)
   +\gamma\varepsilon_{\mu\nu\rho\sigma}
   F_{\mu\nu}(x)F_{\rho\sigma}(x+\widehat\mu+\widehat\nu)
   +\partial_\mu^*k_\mu(x),
\eqn\onexthree
$$
where $\alpha$, $\beta_{\mu\nu}$ and~$\gamma$ are constants and
$\varepsilon_{\mu\nu\rho\sigma}$~is the Levi-Civita symbol;
$F_{\mu\nu}(x)$ is the Abelian field strength and
$\widehat\mu$~stands for the unit vector in direction~$\mu$. In the
last term, the current~$k_\mu(x)$ is a gauge invariant quantity that
locally depends on the gauge field; the symbol
$\partial_\mu^*$~stands for the backward difference operator. In the
next section, we summarize our notation and a set of assumptions under
which Eq.~\onexthree\ holds.

In~Eq.~\onexthree, the gauge invariance of the current~$k_\mu(x)$
distinguishes the last term from other terms which may also be
written as total divergences (see below) but only in terms of gauge
variant currents. Equation~\onexthree\ is important because it tells
us the structure of the axial anomaly with a finite lattice spacing.
(The first two terms are absent for the axial anomaly in four
dimensions, because in this case $q(x)$ behaves as pseudo-scalar
quantity under the lattice symmetries.) Roughly speaking, the
current~$k_\mu(x)$ represents a lattice artifact in the axial anomaly
and it may be removed by a gauge invariant local redefinition of the
axial current. By defining chiral properties with respect to the GW
chiral matrix~[\NAR], the (covariant) gauge anomaly in chiral gauge
theories is also given by the field~$q(x)$. Therefore, the
theorem~\onexthree\ also specifies the structure of the Abelian gauge
anomaly with a finite lattice spacing. The coefficient~$\gamma$ is
given by the coefficient of the anomaly in the continuum
theory.\foot{%
This may be seen as follows: From the dimensional analysis, the
coefficient~$\gamma$ cannot depend on the spacing~$a$. On the other
hand, if the GW-Dirac operator~$D$ behaves properly in the
limit~$a\to0$, it can be shown~[\FUJI] that~$\lim_{a\to0}q(x)=%
\gamma'\varepsilon_{\mu\nu\rho\sigma}F_{\mu\nu}(x)F_{\rho\sigma}(x)$
reproduces the anomaly in the continuum theory. A comparison
with~Eq.~\onexthree\ in the limit~$a\to0$ shows~$\gamma=\gamma'$.}
These facts suggest that, when the anomaly cancellation condition in
the {\it continuum\/} theory is fulfilled, i.e., when $\gamma=0$, a
gauge invariant lattice formulation of Abelian chiral gauge theories
is possible, by removing the artificial breaking of the gauge
symmetry on the lattice, i.e., the current $k_\mu(x)$. In the
meantime, Neuberger~[\NEUB] pointed out that a similar ``anomaly
cancellation'' mechanism works with the overlap formalism~[\NARA], at
least for a particular kind of gauge field configurations. Finally,
on the basis of the observation~\onexthree, L\"uscher~[\LUSCH] gave
an existence proof of a gauge invariant lattice formulation of
Abelian chiral gauge theories, which is consistent with other
physical requirements. See Ref.~[\FUJIK] for recent related works.

In this paper, we present an alternative proof of the
theorem~\onexthree\ with the aim of a more transparent understanding
of it. The idea is the following. The statements~\onextwo\
and~\onexthree\ have the counterpart in the continuum field theory,
which has been proven~[\BRA,\DUB] by utilizing algebraic techniques
introduced in~Ref.~[\ZUM]. It is thus natural to expect similar
algebraic techniques may be applied to the present problem in the
lattice gauge theory. However, as is well-known, the Leibniz rule
does not hold for differences on a lattice and this has been an
obstruction to such an algebraic approach. Here we avoid this
difficulty related to the Leibniz rule by introducing the notion of
non-commutative differential calculus~(NCDC) on discrete set and
groups~[\CON--\DIN]. Some applications of this notion to lattice
(field) theories have been discussed in Refs.~[\DIM,\DIN]. See also
Ref.~[\BAL]. By using NCDC, we may apply the almost identical
cohomological arguments as in the continuum theory, such as the
descent equation~[\ZUM], to the present problem. As an extra bonus of
this algebraic approach, we can not only prove Eq.~\onexthree\ but
also easily obtain its generalization to arbitrary dimensions,
because the algebra involved is independent of the dimensionality of
the lattice.

Here we are not claiming that a certain non-commutativity has to be
implemented on the lattice. We use NCDC simply as a convenient tool
in intermediate steps of extracting relations which are valid on a
lattice even in the conventional sense. It must be possible to show
the resulting relations without relying on NCDC; this is actually the
case as is briefly discussed in Ref.~[\FUJIW].

In a recent interesting paper~[\LUSCHE], it was pointed out that a
solution to~Eq.~\onextwo\ in~$(4{+}2)$-dimensional non-Abelian gauge
theories (here $4$-dimensions are discretized and $2$-dimensions are
continuous) is directly related to a gauge invariant lattice
formulation of non-Abelian chiral gauge theories in $4$-dimensions.
We expect that algebraic approaches similar to the present one should
be useful to study such problems also in non-Abelian theories. This
is our another motivation of the present study and the present paper
may be regarded as a first step toward this direction.

The organization of this paper is as follows. In the next section,
we present some preparation including a set of basic assumptions, a
brief review of NCDC and a quick introduction of the BRST
transformation. In Sec.~3, we state our main result which generalizes
Eq.~\onexthree\ to arbitrary dimensions. To prove our main theorem,
we need several lemmas, and each of these lemmas has the counterpart
in the continuum field theory~[\BRA,\DUB]. In Sec.~4, we give the
proof of these lemmas. Section~5 is devoted to concluding remarks.

\chapter{Preparation}
\section{Basic assumptions}
Throughout this paper, we consider a $D$-dimensional hypercubic
regular lattice with an infinite extent. The lattice spacing is taken
to be unity~$a=1$. We define the forward difference
operator~$\partial_\mu$ and the backward difference
operator~$\partial_\mu^*$ as
$$
\eqalign{
   \partial_\mu f(x)=f(x+\widehat\mu)-f(x),
\cr
   \partial_\mu^* f(x)=f(x)-f(x-\widehat\mu),
\cr
}
\eqn\twoxone
$$
where $f(x)$ is an arbitrary function on the lattice and
$\widehat\mu$ stands for the unit vector in direction~$\mu$. We also
introduce the Abelian field strength~$F_{\mu\nu}(x)$ by
$$
   F_{\mu\nu}(x)={1\over i}\ln U_\mu(x)U_\nu(x+\widehat\mu)
   U_\mu(x+\widehat\nu)^{-1}U_\nu(x)^{-1},
\eqn\twoxtwo
$$
where $U_\mu(x)\in U(1)$~is the link variable on the link that
connects the points~$x$ and~$x+\widehat\mu$. We always assume the
``admissibility'' of the gauge field configuration~[\LUSC]:
$$
   \sup_{x,\mu,\nu}|F_{\mu\nu}(x)|<\epsilon,
\eqn\twoxthree
$$
for a fixed constant $0<\epsilon<\pi/3$. Under this condition, one
can associate (up to integer valued gauge transformations) the gauge
potential~$A_\mu(x)$ with the link variable as~$U_\mu(x)=%
e^{iA_\mu(x)}$, and the field strength is given by~$F_{\mu\nu}(x)=%
\partial_\mu A_\nu(x)-\partial_\nu A_\mu(x)$~[\LUSC]. In particular,
the Bianchi identity~$\partial_{[\mu}F_{\nu\rho]}(x)=0$ holds
under the condition~\twoxthree.

Let $f(x)$~be a functional of the external gauge potential~$A_\mu$.
We say that $f(x)$ locally depends on the gauge potential~$A_\mu$
if the dependence of~$f(x)$ on the gauge potential at distant points
is at least exponentially weak. We say that $f(x)$ smoothly depends
on the gauge potential~$A_\mu$ if the dependence of~$f(x)$ is
sufficiently smooth so that an arbitrary number of derivatives with
respect to~$A_\mu$ around~$A_\mu=0$ may be considered. To prove
Eq.~\onexthree\ and its generalization, the axial
anomaly~$q(x)$~\onexone\ must locally and smoothly depend on the
external gauge potential. From Eq.~\onexone, this is equivalent to the
requirement that so is the Dirac operator~$D(x,x)$. Although a Dirac
operator which satisfies GW relation cannot be ultralocal in
general~[\HOR], i.e., the dependence cannot be restricted within any
finite lattice distance, the dependence on distant points is
exponentially weak~[\HER] at least for the overlap-Dirac
operator~[\NEU]; we assume this holds in general. The necessary
smoothness of the dependence is also ensured by the overlap-Dirac
operator~[\HER,\LUSC,\LUSCH]; we assume this also holds in general.

\section{Non-commutative differential calculus}
Here we introduce the differential calculus on the lattice. See
also~Ref.~[\LUSC]. The bases of the 1-form on a $D$-dimensional
hypercubic lattice are defined as abstract objects which satisfy the
Grassmann algebra:
$$
   dx_1,dx_2,\cdots,dx_D,\qquad dx_\mu dx_\nu=-dx_\nu dx_\mu.
\eqn\twoxfive
$$
By using these bases, a generic $n$-form is defined by
$$
   f(x)={1\over n!}\,f_{\mu_1\cdots\mu_n}(x)
   \,dx_{\mu_1}\cdots dx_{\mu_n},
\eqn\twoxsix
$$
where the summation of repeated indices is understood. The exterior
derivative (or difference more precisely) is defined by the forward
difference operator in~Eq.~\twoxone\ as
$$
   df(x)={1\over n!}\,\partial_\mu f_{\mu_1\cdots\mu_n}(x)
   \,dx_\mu dx_{\mu_1}\cdots dx_{\mu_n}.
\eqn\twoxseven
$$
This definition immediately implies the nilpotency of the exterior
derivative~$d^2=0$. We also introduce the gauge potential 1-form and
the field strength 2-form by
$$
\eqalign{
   &A(x)=A_\mu(x)\,dx_\mu,
\cr
   &F(x)
   ={1\over2}F_{\mu\nu}(x)\,dx_\mu dx_\nu
   ={1\over2}\,[\partial_\mu A_\nu(x)-\partial_\nu A_\mu(x)]
   \,dx_\mu dx_\nu
   =dA(x).
\cr
}
\eqn\twoxeight
$$
Note that the Bianchi identity~$dF(x)=0$ holds, once the field
strength is expressed by the gauge potential in this way.

The essence of NCDC~[\CON,\SIT] on an infinite lattice~[\DIM,\DIN]
is the following relation:
$$
   dx_\mu f(x)=f(x+\widehat\mu)\,dx_\mu.
\eqn\twoxnine
$$
Namely a function on the lattice and the basis of 1-form do not
simply commute; rather the argument of the function is shifted along
$\mu$-direction by one unit when commuting these two objects. The
remarkable fact, which follows from the non-commutativity~\twoxnine,
is that the Leibniz rule of the exterior derivative~$d$ holds.
With~Eqs.~\twoxseven, \twoxone\ and~\twoxnine, one can easily confirm
that
$$
   d\,[f(x)g(x)]=df(x)g(x)+(-1)^nf(x)dg(x),
\eqn\twoxten
$$
for arbitrary forms $f(x)$ and~$g(x)$ (here $f(x)$ is an $n$-form).
The validity of this Leibniz rule is crucial in our approach. We
emphasize again that the exterior derivative~$d$ is defined with the
difference operator~$\partial_\mu$ in~Eq.~\twoxone.

\section{Abelian BRST transformation}
In this subsection, we quickly introduce the BRST transformation and
the associated differential form in the BRST superspace~[\ZUM]. It is
possible to study the present problem in Abelian gauge theories
without introducing such a sophisticated machinery~[\FUJIW]. However,
we think that formulating the problem in this way would be useful to
generalize the present approach to non-Abelian cases. The Abelian BRST
transformation~$\delta_B$ is defined as usual by (but note that
$\partial_\mu$~is the difference operator)
$$
   \delta_BA_\mu(x)=\partial_\mu c(x),\qquad
   \delta_Bc(x)=0,
\eqn\twoxeleven
$$
where $c(x)$ is the Faddeev-Popov ghost associated with the Abelian
gauge transformation. Note that the field strength is BRST invariant
$\delta_BF_{\mu\nu}(x)=0$ and the BRST transformation is
nilpotent~$\delta_B^2=0$. We may also introduce the
anti-ghost~$\overline c(x)$ and the Nakanishi-Lautrup field~$B(x)$,
and set $\delta_B\overline c(x)=iB(x)$ and $\delta_BB(x)=0$. However
these fields play no role in the following analyses. Next we
introduce the Grassmann coordinate~$\theta$ and define the BRST
exterior derivative by
$$
   s=\delta_B\cdot d\theta.
\eqn\twoxtwelve
$$
Note that the usual 1-form~$dx_\mu$ and the BRST 1-form~$d\theta$
anti-commute each other~$dx_\mu d\theta=-d\theta dx_\mu$ and the BRST
1-form~$d\theta$ commutes with itself $d\theta d\theta\neq0$.
Therefore, for a Grassmann-even(-odd) $n$-form~$f(x)$, we have
$$
   d\theta f(x)=\pm(-1)^nf(x)\,d\theta.
\eqn\twoxfourteen
$$
Also we have
$$
   s^2=\{s,d\}=0,
\eqn\twoxthirteen
$$
where the first relation follows from~$\delta_B^2=0$. Finally, we
introduce the ghost 1-form by
$$
   C(x)=c(x)\,d\theta.
\eqn\twoxfifteen
$$
In terms of these forms, the BRST transformation~\twoxeleven\ is
expressed as
$$
   sA(x)=-dC(x),\qquad sC(x)=0.
\eqn\twoxsixteen
$$
Of course we have $sF(x)=0$. Note that $c(x)^2=C(x)^2=0$ in Abelian
cases.

\chapter{The main result}
\noindent
We can now state our main result:
\proclaim Theorem.
Let $q(x)$~be a gauge invariant scalar field on a $D$-dimensional
infinite hypercubic lattice. Suppose that the field~$q(x)$ is locally
and smoothly depending on the Abelian gauge potential~$A_\mu$ and
that $q(x)$~is topological, namely,
$$
   \sum_x\delta q(x)=0,
\eqn\threexone
$$
under an arbitrary local variation of the gauge potential~$A_\mu$.
Then $q(x)$ has the following structure:
$$
\eqalign{
   q(x)&=\alpha
   +\sum_{n=1}^{[D/2]}
   \beta_{\mu_1\nu_1\mu_2\nu_2\cdots\mu_n\nu_n}
   F_{\mu_1\nu_1}(x)
   F_{\mu_2\nu_2}(x+\widehat\mu_1+\widehat\nu_1)\cdots
\cr
   &\qquad\qquad\qquad\qquad\qquad\qquad\times
   F_{\mu_n\nu_n}(x+\widehat\mu_1+\widehat\nu_1+\cdots
    +\widehat\mu_{n-1}+\widehat\nu_{n-1})
\cr
   &\quad+\partial_\mu^*k_\mu(x),
}
\eqn\threextwo
$$
where~$\alpha$ and $\beta_{\mu_1\nu_1\cdots\mu_n\nu_n}$ are constants
which are totally anti-symmetric in indices; in particular,
$\beta_{\mu_1\nu_1\cdots\mu_{D/2}\nu_{D/2}}$ is a constant multiple
of the $D$-dimensional Levi-Civita symbol. The current~$k_\mu(x)$ is
gauge invariant and is locally and smoothly depending on the gauge
potential~$A_\mu$ and~$\left.k_\mu(x)\right|_{A=0}=0$.

The part~$FF\cdots F$ of~Eq.~\threextwo\ corresponds to the Chern
character in the continuum theory~[\ZUM] and the gauge invariant local
current~$k_\mu(x)$ may be regarded as a lattice artifact in the
anomaly. Equation~\threextwo\ reproduces Eq.~\onexthree\ for~$D=4$
($\beta_{\mu\nu\rho\sigma}=\gamma\varepsilon_{\mu\nu\rho\sigma}$) and
provides its higher-dimensional generalization. The coordinate
dependences in the ``Chern character'' $FF\cdots F$
in~Eq.~\threextwo\ can naturally be understood in the framework of
NCDC, as we will see shortly. In particular, in terms of NCDC, it is
quite easy to see that $FF\cdots F$ is in fact a total divergence on
the lattice and it has the topological property~\threexone.

The main theorem~\threextwo\ can be shown straightforwardly, once the
following four lemmas are established. The first two lemmas are
concerning the ``De Rham cohomology'' on the infinite lattice. We
simply quote the first one from~Ref.~[\LUSC]:\foot{%
The Poincar\'e lemma for 1-forms on a two-dimensional lattice has
been proven in~Ref.~[\DIMA] in the context of NCDC.}
\proclaim Poincar\'e lemma.
Let $\eta(x)$ be a $p$-form that decreases at least exponentially at
infinity. If it is $d$-closed $d\eta(x)=0$ for $p<D$ or
$\sum_x\eta(x)=0$ for $p=D$, then one can construct a $(p-1)$-form
$\chi(x)$ such that
$$
   \eta(x)=d\chi(x).
\eqn\addthree
$$
The form~$\chi(x)$ also decreases at least exponentially at infinity.

The construction~[\LUSC] of the form~$\chi(x)$ from the original
form~$\eta(x)$ in~Eq.~\addthree\ requires a certain reference point
other than the point~$x$. If one naively applies the construction to a
certain field~$\eta(x)$, with a particular choice of the reference
point (e.g., the origin), then the translational invariance is broken
and the resulting field~$\chi(x)$ does not locally depend on the
external gauge potential even if the original field~$\eta(x)$ locally
depends on the gauge potential.\foot{%
This point was overlooked in the first version of this paper. We are
grateful to M. L\"uscher for calling this point to our attention.}
Thus, in the proof of~Eq.~\onexthree\ in~Ref.~[\LUSC], the
construction is always applied to ``bi-local'' fields such
as~$\eta(x,y)$, by taking~$y$ or~$x$ as the reference point. This
complication associated to the reference point may however be avoided
as follows:
\proclaim Algebraic Poincar\'e lemma.
Let $\eta(x)$ be a $p$-form that is locally and smoothly depending on
the gauge potential~$A_\mu$ and $\left.\eta(x)\right|_{A=0}=0$. If it
is $d$-closed $d\eta(x)=0$ for~$p<D$ or $\sum_x\eta(x)=0$ for~$p=D$,
then there exists a $(p-1)$-form~$\chi(x)$ such that
$$
   \eta(x)=d\chi(x).
\eqn\threexthree
$$
Moreover the form~$\chi(x)$ is also locally and smoothly depending on
the gauge potential~$A_\mu$ and $\left.\chi(x)\right|_{A=0}=0$.

The above lemma~\threexthree\ is a simple consequence of the original
Poincar\'e lemma~\addthree, as we will see in the next section.
Nevertheless it requires no explicit reference point and its use
considerably simplifies the following arguments; the algebraic
Poincar\'e lemma is rather handy.

The next one is concerning the BRST cohomology in Abelian gauge
theories:
\proclaim Abelian BRST cohomology.
Let $X(x)$ be a form that is locally and smoothly depending on the
gauge potential~$A_\mu$ and possibly on the ghost field~$c$. Suppose
that the form~$X(x)$ has a definite (non-negative) ghost number. Then
there exist the forms $\Omega(x)$ and~$Y(x)$ such that
$$
   sX(x)=0\Leftrightarrow X(x)=
   \cases{
   \Omega(x),&for $X$ with the ghost number zero,\cr
   C(x)\Omega(x)+sY(x),&for $X$ with the ghost number one,\cr
   sY(x),&otherwise,
   }
\eqn\threexfour
$$
where the form $\Omega(x)$ depends on the gauge potential~$A_\mu$ and
it is gauge invariant; the form~$Y(x)$ may depend on the ghost
field~$c$ as well as on the gauge potential. These dependences are
local and smooth.

The next lemma is the crucial one for the proof of the theorem:
\proclaim Abelian covariant Poincar\'e lemma.
Let $\omega_p(x)$ be a gauge invariant $p$-form that is locally and
smoothly depending on the gauge potential~$A_\mu$ and
$\left.\omega_p(x)\right|_{A=0}=0$. If it is
$d$-closed~$d\omega_p(x)=0$ for~$p<D$ or if it is
$d$-exact~$\omega_p(x)=d\chi_{p-1}(x)$ for~$p=D$ with the $(p-1)$-form
$\chi_{p-1}(x)$ that is locally and smoothly depending on the gauge
potential and $\left.\chi_{p-1}(x)\right|_{A=0}=0$, then there exists
a gauge invariant $(p-1)$-form $\eta(x)$ such that
$$
   \omega_p(x)=P(F(x))+d\eta(x),
\eqn\threexfive
$$
where $P(F)$ is a polynomial of the field strength 2-form~$F$
with~$P(0)=0$; all the coefficients of the polynomial are placed on
the right of $F$'s as $P(F)=F^nB+\cdots$. The form $\eta(x)$ is
locally and smoothly depending on the gauge potential~$A_\mu$ and
$\left.\eta(x)\right|_{A=0}=0$.

The proof of the above lemmas (except the first one, for which we
refer to Ref.~[\LUSC]) will be given in the next section. Here we
show how the main theorem~\threextwo\ follows from these lemmas.
(In fact, we need the Poincar\'e lemma~\addthree\ and the Abelian
covariant Poincar\'e lemma~\threexfive\ in the proof of the theorem.
The other two, the algebraic Poincar\'e lemma and the Abelian BRST
cohomology are used to show the Abelian covariant Poincar\'e lemma.)
First, as noted in~Ref.~[\LUSC], an arbitrary smooth functional of
the gauge potential~$A_\mu$ can be represented as
$$
   q(x)=\alpha+\sum_yA_\nu(y)\,j_{\nu,y}(x),
\eqn\threexsix
$$
where $\alpha=\left.q(x)\right|_{A=0}$ and the quantity $j_{\nu,y}(x)$
has been introduced by\foot{%
The substitution~$A_\mu\to tA_\mu$ with $0\le t\le 1$ preserves the
admissibility~\twoxthree\ and thus preserves the locality and the
smoothness of the Dirac operator~[\LUSC].}
$$
   j_{\nu,y}(x)=\int_0^1dt\,
   \left[{\partial q(x)\over\partial A_\nu(y)}\right]_{A\to tA}.
\eqn\addone
$$
We then introduce the $D$-form~$J_{\nu,y}(x)$ by multiplying the
volume form~$d^Dx=dx_1\cdots dx_D$ to~$j_{\nu,y}(x)$,
$J_{\nu,y}(x)=j_{\nu,y}(x)\,d^Dx$. The $x$-integration of the
$D$-form $J_{\nu,y}(x)$ however vanishes due to the topological
property of the field~$q(x)$, Eq.~\threexone:
$$
   \sum_xJ_{\nu,y}(x)
   =\int_0^1dt\,
   \left[\sum_x{\partial q(x)\over\partial A_\nu(y)}\right]_{A\to tA}
   d^Dx=0.
\eqn\threexseven
$$
Also the assumed locality property of~$q(x)$ implies that
$j_{\nu,y}(x)$~\addone\ and~$J_{\nu,y}(x)$ decrease at least
exponentially as $|x-y|\to\infty$. We can thus apply the original
Poincar\'e lemma~\addthree\ to the $D$-form~$J_{\nu,y}(x)$ to yield
$$
   J_{\nu,y}(x)=dT_{\nu,y}(x).
\eqn\addtwo
$$
The important point to note here that the construction~[\LUSC] of
the $(D-1)$-form~$T_{\nu,y}(x)$ guarantees that, by taking the
point~$y$ as the reference point, $T_{\nu,y}(x)$ also possesses the
same locality property as~$J_{\nu,y}(x)$. The smoothness is also
preserved by the construction. We notice that the construction
in~Ref.~[\LUSC] works even with NCDC. Therefore, from Eqs.~\threexsix\
and~\addtwo, we have
$$
\eqalign{
   q(x)\,d^Dx&=\alpha\,d^Dx+d\sum_yA_\nu(y)\,T_{\nu,y}(x)
\cr
   &=\alpha\,d^Dx+d\chi(x).
\cr
}
\eqn\threexeight
$$
Due to the above locality property of~$T_{\nu,y}(x)$, the summation
over~$y$ in the second term is well-convergent and we see that the
$(D-1)$-form~$\chi(x)$ is locally and smoothly depending on the gauge
potential. We also note that the $D$-form~$d\chi(x)$ is gauge
invariant because $q(x)$~is gauge invariant by assumption and
$\left.\chi(x)\right|_{A=0}=0$. From these facts, we see that the
requisites of the Abelian covariant Poincar\'e lemma~\threexfive\ are
fulfilled for the $d\chi$-part of~Eq.~\threexeight\ and thus
$$
   q(x)\,d^Dx=\alpha\,d^Dx+P(F(x))+d\eta(x).
\eqn\threexten
$$
The lemma~\threexfive\ also says that the $(D-1)$-form $\eta(x)$ is
gauge invariant and is locally and smoothly depending on the gauge
potential~$A_\mu$ and $\left.\eta(x)\right|_{A=0}=0$.

To extract the information on the field~$q(x)$ from~Eq.~\threexten,
we have to factor out the volume form~$d^Dx$ from both sides of the
equation. For the monomial of the field strength 2-form, this
operation yields
$$
\eqalign{
   F(x)^nB
   &={1\over2}F_{\mu_1\nu_1}(x)\,dx_{\mu_1}dx_{\nu_1}
   {1\over2}F_{\mu_2\nu_2}(x)\,dx_{\mu_2}dx_{\nu_2}\cdots
   {1\over2}F_{\mu_n\nu_n}(x)\,dx_{\mu_n}dx_{\nu_n}
\cr
   &\qquad\qquad\qquad\qquad\qquad\qquad\times
   {1\over(D-2n)!}\,B_{\rho_1\cdots\rho_{D-2n}}
   dx_{\rho_1}\cdots dx_{\rho_{D-2n}}
\cr
   &={1\over2^n(D-2n)!}\,
   \varepsilon_{\mu_1\nu_1\mu_2\nu_2\cdots\mu_n\nu_n
   \rho_1\cdots\rho_{D-2n}}
   B_{\rho_1\cdots\rho_{D-2n}}
\cr
   &\qquad\qquad\qquad\quad\times
   F_{\mu_1\nu_1}(x)
   F_{\mu_2\nu_2}(x+\widehat\mu_1+\widehat\nu_1)\cdots
\cr
   &\qquad\qquad\qquad\qquad\times
   F_{\mu_n\nu_n}(x+\widehat\mu_1+\widehat\nu_1+\cdots
    +\widehat\mu_{n-1}+\widehat\nu_{n-1})\,d^Dx,
\cr
}
\eqn\threexeleven
$$
where $\varepsilon_{\mu_1\cdots\mu_D}$ is the $D$-dimensional
Levi-Civita symbol (we define $\varepsilon_{12\cdots D}=1$). In
deriving the last expression, we have used the basic property of NCDC,
Eq.~\twoxnine. By defining $\beta$ as the dual of~$B$, we thus obtain
the $FF\cdots F$ part of~Eq.~\threextwo. Note that the coordinate
dependences of the ``Chern character'' have been easily obtained by
the repeated applications of the non-commutative rule~\twoxnine.
Finally the $d$-exact piece of~Eq.~\threexten, $d\eta(x)$, can be
expressed as a total divergence of the dual vector~$k_\mu(x)$,
$k_\mu(x)=%
\varepsilon_{\mu\nu_1\cdots\nu_{D-1}}%
\eta_{\nu_1\cdots\nu_{D-1}}(x+\widehat\mu)/(D-1)!$, times the volume
form~$d^Dx$. This completes the proof of the theorem~\threextwo.

In the context of NCDC, the topological nature of the Chern character
is also manifest because we can use the Leibniz rule. For example, we
have
$$
\eqalign{
   F(x)^{D/2}&=d\,[A(x)F(x)^{D/2-1}]
\cr
   &={1\over2^{D/2-1}}\,
   \varepsilon_{\mu_1\nu_1\mu_2\nu_2\cdots\mu_{D/2}\nu_{D/2}}
\cr
   &\quad\times
   \partial_{\mu_1}\Bigl[
   A_{\nu_1}(x)
   F_{\mu_2\nu_2}(x+\widehat\nu_1)
   F_{\mu_3\nu_3}(x+\widehat\nu_1+\widehat\mu_2+\widehat\nu_2)\cdots
\cr
   &\qquad\qquad\times
   F_{\mu_{D/2}\nu_{D/2}}(x+\widehat\nu_1
   +\widehat\mu_2+\widehat\nu_2+\cdots
    +\widehat\mu_{D/2-1}+\widehat\nu_{D/2-1})\Bigr]\,d^Dx,
\cr
}
\eqn\threextwelve
$$
where uses of the Bianchi identity~$dF(x)=0$ and the basic property of
NCDC~\twoxnine\ have been made. In~Eqs.~\threexeleven\
and~\threextwelve, we {\it must\/} use the non-commutative
rule~\twoxnine, because the lemma~\threexfive\ holds only in terms of
NCDC, as we will see in the next section. In other words,
Eq.~\threexten\ is meaningful only in the context of NCDC. However,
relations among the coefficient functions such as Eq.~\threextwo\
hold irrespective of use of NCDC. As is assuring, the
theorem~\threextwo\ can also be proven~[\FUJIW] by a direct extension
of the argument of~Ref.~[\LUSC].

\chapter{Proof of the lemmas}
In this section, we present the detailed proof of the lemmas in the
preceding section.
\section{Algebraic Poincar\'e lemma}
To show Eq.~\threexthree, we start with the representation
$$
   \eta(x)=\sum_yA_\nu(y)J_{\nu,y}(x),
\eqn\addfour
$$
where the $p$-form $J_{\nu,y}(x)$ is given by\foot{%
In the context of NCDC, the argument of a function changes depending
on the order of form bases and the function. For this reason, the
derivative with respect to~$A_\nu(y)$ in this equation must be defined
as the left derivative.}
$$
   J_{\nu,y}(x)=\int_0^1dt
   \left[{\partial\eta(x)\over\partial A_\nu(y)}\right]_{A\to tA},
\eqn\addfive
$$
and we have used the fact that $\left.\eta(x)\right|_{A=0}=0$. From
Eq.~\addfive, it is obvious that $dJ_{\nu,y}(x)=0$ if $d\eta(x)=0$,
and $\sum_xJ_{\nu,y}(x)=0$ if $\sum_x\eta(x)=0$. Also the locality
of~$\eta(x)$ implies that $J_{\nu,y}(x)$ decreases at least
exponentially as $|x-y|\to\infty$. Therefore the form~$J_{\nu,y}(x)$
satisfies the requisite of the original Poincar\'e lemma~\addthree.
We thus have
$$
   J_{\nu,y}(x)=dT_{\nu,y}(x).
\eqn\addsix
$$
As noted in Eq.~\addtwo\ below, the $(p-1)$-form~$T_{\nu,y}(x)$ shares
the same locality property as~$J_{\nu,y}(x)$. Then going back to
Eq.~\addfour, we have
$$
\eqalign{
   \eta(x)&=d\sum_yA_\nu(y)T_{\nu,y}(x)
\cr
   &=d\chi(x).
\cr
}
\eqn\addseven
$$
{}From the locality property of the $(p-1)$-form~$T_{\nu,y}(x)$, it is
obvious that the summation over~$y$ is well-convergent and the
$(p-1)$-form $\chi(x)$ is locally and smoothly depending on the gauge
potential. We also note that $\left.\chi(x)\right|_{A=0}=0$. In this
way, we have the algebraic Poincar\'e lemma, Eq.~\threexthree.

\section{Abelian BRST cohomology}
The first step to show the BRST cohomology~\threexfour\ is to choose
a convenient set of bases in the functional space on the lattice, that
is analogous to the jet variables in the continuum theory~[\BRA].
Noting the relations $f(x+\widehat\mu)=(1+\partial_\mu)f(x)$ and
$f(x-\widehat\mu)=(1-\partial_\mu^*)f(x)$, we see that the gauge
potential at an arbitrary point can be expressed as a linear
combination of the terms with a structure
$$
   (\partial_1)^{m_1}(\partial_2)^{m_2}\cdots(\partial_D)^{m_D}
   A_\mu(x),
\eqn\fourxone
$$
where $x$~is a certain reference point and $m_\mu$'s are integer. In
Eq.~\fourxone, we have introduced the notation
$$
   (\partial_\mu)^m=\cases{\partial_\mu^m&for $m>0$,\cr
                           \partial_\mu^{*-m}&for $m<0$.\cr}
\eqn\fourxtwo
$$
Note that, in~Eq.~\fourxone, {\it either\/} the forward {\it or\/}
the backward difference operator appears in each coordinate direction,
but not both. Obviously the variables~\fourxone\ with a different set
of $m$'s are linearly independent, because otherwise there exists a
linear relation among the gauge potentials~$A_\mu$ at different
points. To discuss the BRST cohomology, however, it is convenient to
separate gauge invariant combinations from the set of variables
of~Eq.~\fourxone. To do this, we note the following relations which
``exchange'' the order of indices of the difference operator and of
the gauge potential
$$
\eqalign{
   &\partial_\mu A_\nu(x)=\partial_\nu A_\mu(x)+F_{\mu\nu}(x),
\cr
   &\partial_\mu^*A_\nu(x)=(1-\partial_\mu^*)\partial_\nu A_\mu(x)
   +(1-\partial_\mu^*)F_{\mu\nu}(x).
\cr
}
\eqn\fourxthree
$$
Namely, by a repeated use of these identities, we can always make the
indices of the difference operators in~Eq.~\fourxone\ less or equal
to that of the gauge potential~$A_\mu$, up to terms linear in the
field strength~$F_{\mu\nu}$. After this manipulation, we can apply
the identity
$$
   \partial_\mu^*\partial_\mu=\partial_\mu-\partial_\mu^*,
\eqn\fourxfour
$$
to the product of the forward and the backward difference operators
in the {\it same\/} direction, if necessary. In this way, the
variable~\fourxone\ can be expressed as a linear combination of
$$
   A_i=(\partial_1)^{m_1}\cdots(\partial_\mu)^{m_\mu}A_\mu(x)\quad
   {\rm and}\quad
   F_i=(\partial_1)^{m_1}\cdots(\partial_D)^{m_D}F_{\mu\nu}(x).
\eqn\fourxfive
$$
It is also obvious that the variables $\{A_i\}$ and $\{F_i\}$ are
linearly independent. For variables for the ghost field~$c$, we choose
$$
   c_i=\delta_BA_i
   =(\partial_1)^{m_1}\cdots(\partial_\mu)^{m_\mu}\partial_\mu c(x).
\eqn\fourxsix
$$
Note that the combination of the difference operators is somewhat
different from that of the representation~\fourxone, because
Eq.~\fourxsix\ always contains at least one forward difference
operator as $\cdots\partial_\mu c(x)$ and a single backward difference
$\partial_\mu^*c(x)$ for example does not appear in~Eq.~\fourxsix.
However the latter can be expressed as $\partial_\mu^*c=
\partial_\mu c-\partial_\mu^*\partial_\mu c$ by the
identity~\fourxfour. The set of variables of Eq.~\fourxsix\ added
with~$c(x)$ is, therefore, equivalent to the
choice~$\{(\partial_1)^{m_1}\cdots(\partial_D)^{m_D}c(x)\}$. The
variables $\{c_i\}$ and $c(x)$ span a complete set for the ghost
field.

With these preparations, we now turn to the proof of Eq.~\threexfour.
We assume that the function~$X$ in~Eq.~\threexfour\ is expressed in
terms of the variables $\{A_i\}$, $\{F_i\}$, $\{c_i\}$ and $c(x)$. To
eliminate the ambiguity associated to the non-commutativity, we also
assume that all the form bases are placed on the right of the field
variables. We next note that the BRST transformation can be expressed
in terms of these variables as
$$
   \delta_B=\sum_ic_i{\partial\over\partial A_i},
\eqn\fouexseven
$$
because the variables $\{F_i\}$, $\{c_i\}$ and $c(x)$ are BRST
invariant. We also introduce the Grassmann-odd operator~$r$
$$
   r=\sum_iA_i{\partial\over\partial c_i}.
\eqn\fourxeight
$$
Then the anti-commutator of~$\delta_B$ and $r$ gives rise to the
number operator of $\{A_i\}$ and~$\{c_i\}$:
$$
   \{\delta_B,r\}
   =\sum_i\left(A_i{\partial\over\partial A_i}
                +c_i{\partial\over\partial c_i}\right).
\eqn\fourxnine
$$
Finally, we introduce the auxiliary parameter~$t$ in the
function~$X$ by rescaling the variables $\{A_i\}$ and~$\{c_i\}$ as
$A_i\to tA_i$ and $c_i\to tc_i$ (the variables $\{F_i\}$ and $c(x)$
are {\it not\/} rescaled). Then by writing
$X_t=\left.X\right|_{A_i\to tA_i,c_i\to tc_i}$, we have
$$
\eqalign{
   X&=X_1=X_0+\int_0^1dt\,{\partial X_t\over\partial t}
\cr
   &=X_0+\int_0^1{dt\over t}\,\{\delta_B,r\}X_t,
\cr
}
\eqn\fourxten
$$
where we have used Eq.~\fourxnine. Since $\delta_B(tA_i)=tc_i$, if
the function~$X$ is BRST invariant $\delta_BX=0$, then the rescaled
one~$X_t$ is also BRST invariant $\delta_BX_t=0$. Therefore, from
Eq.~\fourxten, we have
$$
   \delta_BX=0\Rightarrow
   X=X_0+\delta_B\int_0^1{dt\over t}\,rX_t,
\eqn\fourxeleven
$$
The first term~$X_0$, which depends only on the gauge invariant
variables $\{F_i\}$ and $c(x)$, is the BRST non-trivial piece (note
that the BRST transformation of a certain function is proportional
to~$c_i$) and the second term gives the BRST exact piece.\foot{%
The integral over the parameter~$t$ is converging at $t=0$
because the possible $O(t^0)$~part of $X_t$ is eliminated by the
derivative $\partial/\partial c_i$ in~$r$.}
However, since $c(x)^n=0$ for~$n\geq2$ in Abelian gauge theories, we
have the non-trivial piece only when the ghost number of the form~$X$
is zero or one. This shows (after supplementing an appropriate
$d\theta$) the Abelian BRST cohomology~\threexfour. The locality and
the smoothness of the forms $\Omega$ and~$Y$ in~Eq.~\threexfour\ are
obvious from the representation~\fourxeleven\ and the locality and
the smoothness of the form~$X$.

\section{Abelian covariant Poincar\'e lemma}
In the context of NCDC, the lemma~\threexfive\ can be proven in an
almost identical way as in the continuum theory~[\BRA], because we
can use the Leibniz rule. To illustrate this point, we will emphasize
how the Leibniz rule is used in the following proof. We show
Eq.~\threexfive\ by mathematical induction. For $p=0$, the lemma
trivially holds by choosing~$P=0$ because of the algebraic Poincar\'e
lemma~\threexthree. Let us assume the lemma holds for $p=0$, $1$,
$\cdots$, $n-1$. Now, for $n<D$, $\omega_n$ is $d$-closed and thus is
$d$-exact by the algebraic Poincar\'e lemma. For~$n=D$, it is
$d$-exact by assumption. Also $\omega_n$~is $s$-closed because
$\omega_n$~is gauge invariant:
$$
   \omega_n=d\chi_{n-1}^0,\qquad s\omega_n=0,
\eqn\fourxsix
$$
where the upper index of~$\chi$ stands for the ghost number and the
lower index stands for the degree of the form. Note that the
$(n-1)$-form~$\chi_{n-1}^0$ is locally and smoothly depending on the
gauge potential~$A_\mu$ and $\left.\chi_{n-1}^0\right|_{A=0}=0$, as
the result of the algebraic Poincar\'e lemma.

The two conditions~\fourxsix\ lead to the descent equation for the
sequence of~$\chi$. Namely, since $0=s\omega_n=sd\chi_{n-1}^0=%
- -ds\chi_{n-1}^0$, the algebraic Poincar\'e lemma states that\foot{%
The algebraic Poincar\'e lemma in its original form~\threexthree\
applies to a form that depends only on the gauge potential~$A_\mu$.
However it is straightforward to generalize the lemma for forms which
contain the ghost field~$c$ as well. The
property~$\left.\chi(x)\right|_{A=0}=0$ is then replaced
by~$\left.\chi(x)\right|_{A=c=0}=0$.}
$s\chi_{n-1}^0=d\chi_{n-2}^1$. Similarly since $0=s^2\chi_{n-1}^0=%
sd\chi_{n-2}^1=-ds\chi_{n-2}^1$, we have $s\chi_{n-2}^1=d\chi_{n-3}^2$
again by the algebraic Poincar\'e lemma. Repeating this procedure, we
have
$$
\eqalign{
   s\chi_{n-1-g}^g&=d\chi_{n-2-g}^{g+1},\qquad0\leq g<n-1,
\cr
   s\chi_0^{n-1}&=0.
}
\eqn\fourxseven
$$
Note that all $\chi$'s are locally and smoothly depending on the gauge
potential and on the ghost field and that
$\left.\chi\right|_{A=c=0}=0$ by the algebraic Poincar\'e lemma.

Now the solution of the last equation in Eq.~\fourxseven\
with~$n\neq1$ is given by the Abelian BRST cohomology~\threexfour,
$$
   \chi_0^{n-1}=\cases{
   C\Omega_0+sb_0,&for $n=2$,\cr
   sb_0,& otherwise,}
\eqn\fourxeight
$$
where the 0-form~$\Omega_0$ is gauge invariant and $\Omega_0$
and~$b_0$ are locally and smoothly depending on the gauge
potential and on the ghost field. But by redefining $\chi_0^{n-1}$
and~$\chi_1^{n-2}$ as
$$
   \chi_0^{n-1}\to\chi_0^{n-1}+sb_0,\qquad
   \chi_1^{n-2}\to\chi_1^{n-2}-db_0,
\eqn\fourxnine
$$
we can completely eliminate $b_0$-dependences from the descent
equation~\fourxseven. Note that this redefinition preserves the
property~$\left.\chi\right|_{A=c=0}=0$, because $sb_0$ and~$db_0$
cannot be a constant. Without loss of generality, therefore, we can
take
$$
   \chi_0^{n-1}=\cases{
   C\Omega_0,&for $n=2$,\cr
   0,& otherwise.}
\eqn\fourxten
$$
Repeating this argument in Eq.~\fourxseven, we see that the
non-trivial part of the descent equation is given by
$$
\eqalign{
   s\chi_{n-1}^0&=d\chi_{n-2}^1,
\cr
   s\chi_{n-2}^1&=0.
\cr
}
\eqn\fourxeleven
$$

The solution of the last equation in Eq.~\fourxeleven\ can be taken
as~$\chi_{n-2}^1=C\Omega_{n-2}=%
C(B_{n-2}+\omega_{n-2})$, here $B_{n-2}$ stands for the constant part
of~$\Omega_{n-2}$, $\left.\Omega_{n-2}\right|_{A=0}=B_{n-2}$
and~$\left.\omega_{n-2}\right|_{A=0}=0$. The form~$\omega_{n-2}$ is
locally and smoothly depending on the gauge potential~$A_\mu$ and is
gauge invariant. The first relation in~Eq.~\fourxeleven\ then yields
$$
\eqalign{
   s\chi_{n-1}^0&=d[C(B_{n-2}+\omega_{n-2})]
\cr
   &=dC(B_{n-2}+\omega_{n-2})-Cd\omega_{n-2}
\cr
   &=-sA(B_{n-2}+\omega_{n-2})-Cd\omega_{n-2},
\cr
}
\eqn\fourxtwelve
$$
where we have used the Leibniz rule and~Eq.~\twoxsixteen. Now the
relation~\fourxtwelve\ holds for arbitrary configurations of the
ghost field~$c(x)$. When the ghost field is a constant~$c(x)=c$,
Eq.~\fourxtwelve\ reduces to $Cd\omega_{n-2}(x)=0$ because other
terms in the equation are proportional to the difference of the ghost
field~$\partial_\mu c(x)=c(x+\widehat\mu)-c(x)$ etc. Therefore we see
that the $(n-2)$-form~$\omega_{n-2}$ must be $d$-closed and
$$
   s[\chi_{n-1}^0+A(B_{n-2}+\omega_{n-2})]=0,\qquad
   d\omega_{n-2}=0.
\eqn\fourxthirteen
$$
The solution of the first equation in Eq.~\fourxthirteen\ is given by
the BRST cohomology~\threexfour,
$$
   \chi_{n-1}^0=-A(B_{n-2}+\omega_{n-2})+\Omega_{n-1}.
\eqn\fourxfourteen
$$
Note that the $(n-1)$-form~$\Omega_{n-1}$ is gauge invariant and
again is locally and smoothly depending on~$A_\mu$ and
$\left.\Omega_{n-1}\right|_{A=0}=0$.

Finally, substituting Eq.~\fourxfourteen\ in~Eq.~\fourxsix, we have
$$
\eqalign{
   \omega_n&=d[-A(B_{n-2}+\omega_{n-2})+\Omega_{n-1}]
\cr
   &=-F(B_{n-2}+\omega_{n-2})+Ad\omega_{n-2}+d\Omega_{n-1}
\cr
   &=-F(B_{n-2}+\omega_{n-2})+d\Omega_{n-1}.
\cr
}
\eqn\fourxfifteen
$$
Here we have used the Leibniz rule and the fact that the
form~$\omega_{n-2}$ is $d$-closed $d\omega_{n-2}=0$, as is shown in
Eq.~\fourxthirteen. However, by the induction hypothesis, the
lemma~\threexfive\ applies to the gauge invariant closed
$(n-2)$-form~$\omega_{n-2}$ (recall that
$\left.\omega_{n-2}\right|_{A=0}=0$):
$$
   \omega_{n-2}=P(F)+d\eta.
\eqn\fourxsixteen
$$
Equations~\fourxfifteen\ and~\fourxsixteen\ then show that the
lemma~\threexfive\ holds for~$\omega_n$ also:
$$
\eqalign{
   \omega_n&=-F[B_{n-2}+P(F)]-Fd\eta+d\Omega_{n-1}
\cr
   &=-F[B_{n-2}+P(F)]+d(-F\eta+\Omega_{n-1}),
\cr
}
\eqn\fourxseventeen
$$
where uses of the Leibniz rule and the Bianchi identity~$dF=0$ have
been made. Note that all the coefficients in~$P(F)$ appear on the
right of~$F$'s in the iteration of Eqs.~\fourxsixteen\
and~\fourxseventeen. The locality and the smoothness of the gauge
invariant quantity in the round brackets in the second equation in
Eq.~\fourxseventeen\ are obvious from the above argument. In this way,
we establish the lemma~Eq.~\threexfive. 

\chapter{Conclusion}
In this paper, we have presented an algebraic proof of the
theorem~\onexthree\ and its generalization to arbitrary
dimensions~\threextwo\ on the basis of cohomological arguments. The
basic tool in our approach is NCDC which makes the Leibniz rule of
exterior derivatives valid on a lattice. By using NCDC, we can readily
transcribe the proof of the corresponding statement in the continuum
theory, such as the (Abelian) linearized covariant Poincar\'e
lemma~[\BRA], into the language of lattice theory. We can also easily
understand the topological nature of the Chern character on a
lattice. In these respects, our algebraic approach is complementary
to the constructive proof in~Ref.~[\LUSC].

The explicit form of the gauge invariant local current~$k_\mu(x)$
in~Eqs.~\onexthree\ and~\threextwo\ is of great interest because it
is a crucial ingredient in the Abelian lattice chiral gauge theories
formulated in~Ref.~[\LUSCH]. For example, the gauge invariant
effective action on an infinite lattice is given by a simple integral
representation, once the explicit form of the current~$k_\mu(x)$ is
known~[\SUZ]. From Eq.~\threexten, we see that the current~$k_\mu(x)$
or equivalently the form~$\eta(x)$ can be constructed by applying the
algebraic Poincar\'e lemma~\threexthree\ to
$[q(x)-\alpha]\,d^Dx-P(F(x))$. From the proof of the algebraic
Poincar\'e lemma in~Sec.~4.1 (and from the proof of the original
Poincar\'e lemma~[\LUSC]), this implies that the current~$k_\mu(x)$
is given by a linear combination of a product of the gauge
potential~$A_\mu$ and the first derivative of~$q(x)$ with respect
to~$A_\mu$. Unfortunately, however, the resulting expression seems to
be too complicated to be of any practical use.

As a simple application of our theorem~\threextwo, we may consider
the general gauge invariant solution
of~$\sum_x\allowbreak\int dt\,ds\,\delta q(z)=0$ in Abelian gauge
theories in the $(4{+}2)$-dimensional space, here $4$-dimensions are
discrete (with coordinates specified by~$x$) and the remaining
$2$-dimensions are continuous (their coordinates are $t$ and~$s$).
This is the cohomological problem posed in~Ref.~[\LUSCHE], in
connection with the gauge invariant formulation of lattice chiral
gauge theories. When the gauge group is Abelian, the general solution
may simply be obtained by taking the classical continuum limit
of~Eq.~\threextwo\ with respect to the continuous $2$-dimensional
space. Then the topological quantity~$q(z)$, up to terms irrelevant
for the discussions in~Ref.~[\LUSCHE], is given by
$$
\eqalign{
   q(z)&=\gamma\varepsilon_{IJKLMN}
   F_{IJ}(x,t,s)
   F_{KL}(x+\widehat I+\widehat J,t,s)
   F_{MN}(x+\widehat I+\widehat J+\widehat K+\widehat L,t,s)
\cr
   &\quad+\partial_\mu^*k_\mu(z)+\partial_t k_t(z)+\partial_sk_s(z),
\cr
}
\eqn\fivexone
$$
where the Latin indices run from 1 to~6 and the Greek indices run
over only from~1 to~4; the vectors along the continuous directions
are zero $\widehat5=\widehat6=0$. For the topological
quantity~$q(z)$ which are relevant for the analysis
in~Ref.~[\LUSCHE], the coefficient~$\gamma$ is proportional to a sum
of the cubic of $U(1)$ charges of Weyl fermions, i.e., the gauge
anomaly. Therefore, when the $U(1)$~gauge anomaly cancellation
condition is fulfilled, the quantity~$q(z)$ is given by total
divergences. The discussion of~Ref.~[\LUSCHE] then shows that it is
possible to formulate Abelian gauge theories on the lattice, while
keeping the exact gauge invariance. Of course this should be so from
the result obtained in~Ref.~[\LUSCH], but the present example clearly
illustrates the usefulness of our result~\threextwo\ for a
higher-dimensional lattice.

We are deeply indebted to M.~L\"uscher for helpful comments on the
previous versions of this paper. We are grateful to
F.~M\"uller-Hoissen for information and to K.~Fujikawa for
helpful comments. K.W. is very grateful to F.~Sakata for the warm
hospitality extended to him and to Faculty of Science of Ibaraki
University for the financial support during his visit at Ibaraki
University.

\bigskip
\titlestyle{\bf Note added:}
The statement made in Eq.~\fourxfive\ below that the variables
$\{F_i\}$ are linearly independent is not correct. The Bianchi
identity and its differences are linear relations among~$F_i$'s.
However, the proof of the BRST cohomology does not change if we
assume that the function~$X$ is expressed in terms of $\{A_i\}$,
$\{c_i\}$, $c(x)$, and the independent set of variables
in~$\{F_i\}$.

\refout

\bye